\documentclass[aps,prx,twocolumn,superscriptaddress,showpacs,floatfix]{revtex4-1}
\usepackage{amsmath,amssymb,graphicx}
\usepackage{wasysym}
\usepackage[utf8]{inputenc}
\usepackage[T1]{fontenc}
\usepackage{xcolor}
\allowdisplaybreaks
\usepackage{rotating} 
\usepackage{bibentry}

\usepackage{textcomp}

\usepackage{bm}

\IfFileExists{siunitx.sty}{\usepackage{booktabs,siunitx}}{}

\usepackage{color}
\definecolor{LinkColor}{rgb}{0.256,0.439,0.588}
\usepackage{hyperref}

\renewcommand{\vec}[1]{\mathbf{#1}}

\usepackage{pifont}

\newcommand{\La}{\line (1,0  ){12}}
\newcommand{\Lb}{\line (3,5 ){6}}

\newcommand{\Ld}{\line (-1,0){12}}
\newcommand{\Le}{\line (-3,-5){6}}

\newcommand{\C} {\circle*{4}}

\newcommand{\LaT}{\rule[-1pt]{0.4cm}{0.2em}}  
\newcommand{\LdT}{\rule[-1pt]{0.4cm}{0.2em}}  
\newcommand{\LbT}{\rotatebox{60}{\rule[-1pt]{0.4cm}{0.2em}}}  
\newcommand{\LeT}{\rotatebox{60}{\rule[-1pt]{0.4cm}{0.2em}}}  

\newcommand{\pA}{\put(-6,-10)}
\newcommand{\pB}{\put(6,-10)}
\newcommand{\pC}{\put(12,0)}

\newcommand{\pZ}{\put(0,0)}

\newcommand{\pAT}{\put(-6.8,-10)} 
\newcommand{\pBT}{\put(5.2,-10)}  

\newcommand{\rhomb}{
  \pA{\C}\pB{\C}\pZ{\C}\pC{\C}
 }

\newcommand{\rhombH}{
  \begin{picture}(22,10)(-8,-6)
    \pA{\LaT}\pB{\Lb}\pZ{\Le}\pZ{\LdT}
    \rhomb
  \end{picture}
}
\newcommand{\rhombV}{
  \begin{picture}(22,10)(-8,-6)
   \pA{\La}\pBT{\LbT}\pAT{\LeT}\pC{\Ld}
    \rhomb
  \end{picture}
}

\newcommand{\ignore}[1]{}
\newcommand{\nobibentry}[1]{{\let\nocite\ignore\bibentry{#1}}}
\newcommand{\bibfnamefont}[1]{#1}
\newcommand{\bibnamefont}[1]{#1}

\begin{document}

\title{Cubic* criticality emerging from a quantum loop model on triangular lattice}

\author{Xiaoxue Ran}
\affiliation{Department of Physics and HKU-UCAS Joint Institute of Theoretical and Computational Physics,The University of Hong Kong, Pokfulam Road, Hong Kong SAR, China}

\author{Zheng Yan}
\affiliation{Department of Physics, School of Science and Research Center for Industries of the Future, Westlake University, Hangzhou 310030,  China}
\affiliation{Institute of Natural Sciences, Westlake Institute for Advanced Study, Hangzhou 310024, China}

\author{Yan-Cheng Wang}
\affiliation{Hangzhou International Innovation Institute, Beihang University, Hangzhou 311115, China}

\author{Junchen Rong}
\affiliation{Institut des Hautes Études Scientifiques, 91440 Bures-sur-Yvette, France}

\author{Yang Qi}
\affiliation{State Key Laboratory of Surface Physics, Fudan University, Shanghai 200433, China}
\affiliation{Center for Field Theory and Particle Physics, Department of Physics, Fudan University,	Shanghai 200433, China}
\affiliation{Collaborative Innovation Center of Advanced Microstructures, Nanjing 210093, China}

\author{Zi Yang Meng}
\email{zymeng@hku.hk}
\affiliation{Department of Physics and HKU-UCAS Joint Institute of Theoretical and Computational Physics,The University of Hong Kong, Pokfulam Road, Hong Kong SAR, China}

\begin{abstract}
Quantum loop and dimer models are archetypal examples of correlated systems with local constraints. Obtaining generic solutions for these models is difficult due to the lack of controlled methods to solve them in the thermodynamic limit. Nevertheless, these solutions are of immediate relevance to both statistical and quantum field theories, as well as the rapidly growing experiments in Rydberg atom arrays and quantum moiré materials, where the interplay between correlation and local constraints gives rise to a plethora of novel phenomena. In a recent work [X. Ran, Z. Yan, Y.-C. Wang, et al, arXiv:2205.04472 (2022)], it was found through sweeping cluster quantum Monte Carlo (QMC) simulations and field theory analysis that the triangular lattice quantum loop model (QLM) hosts a rich ground state phase diagram with lattice nematic, vison plaquette (VP) crystals, and the $\mathbb{Z}_2$ quantum spin liquid (QSL) close to the Rokhsar-Kivelson point. Here, we focus on the continuous quantum critical point separating the VP and QSL phases and demonstrate via both static and dynamic probes in QMC simulations that this transition is of the (2+1)D cubic* universality. In this transition, the fractionalized visons in QSL condense to give rise to the crystalline VP phase, while leaving their trace in the anomalously large anomalous dimension exponent and pronounced continua in the dimer and vison spectra compared with those at the conventional cubic or O(3) quantum critical points. 
\end{abstract}

\date{\today}
\maketitle

\begin{figure}[htp!]
\centering
\includegraphics[width=1\columnwidth]{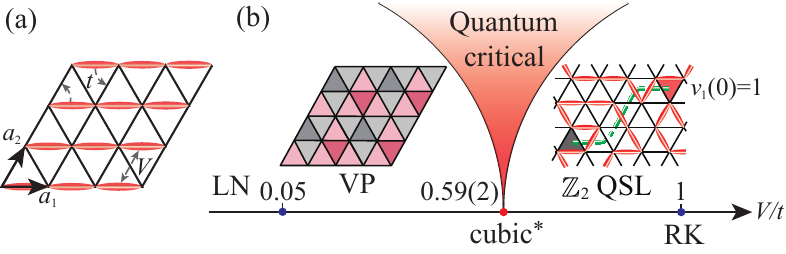}
\caption{{\textbf{Quantum loop model on the triangular lattice and its phase diagram.}} (a) $\mathbf{a}_{1}$ and $\mathbf{a}_{2}$ are the triangular lattice primitive vectors. The $t$ and $V$ terms are the kinetic and potential terms in the Hamiltonian Eq.~\eqref{eq:eq1}, respectively. (b) The transition between the LN and VP crystals is first order~\cite{yanFully2022}, while the transition between VP and the $\mathbb{Z}_{2}$ QSL is continuous and of (2+1)D cubic* universality~\cite{yanFully2022}. The correlation functions of the VP and vison order parameters around the cubic* QCP ($V_c=0.59(2)$) are shown in Figs.~\ref{fig:tcor} and \ref{fig:vcor}. The schematic plot in the VP phase is the real-space vison correlations with the red (grey) color conveying its positive (negative) value in each triangle. The darker color stands for larger absolute values of the correlation functions. The schematic plot in the QSL phase shows two visons connected by a string (the green dashed line), which represents the path $P$ of the vison-vison correlation function $ v_{\gamma}(0) v_{\gamma}(r)$\,$=$\,$(-1)^{N_{P}}$, with $N_{P}$ the number of dimers cut along $P$. Here we set the vison in the lower triangle $v_1(0)=1$ as the reference to fix the gauge.}
\label{fig:model}
\end{figure}

\noindent{\textcolor{blue}{\it Introduction.}---} 
Recently, the ground state phase diagram of the quantum loop model (QLM) on a triangular lattice~\cite{Moessner2001l,moessnerShort2001,Plat2015,Krishanu2015} has been mapped out using the sweeping cluster quantum Monte Carlo (QMC) algorithm~\cite{yanFully2022,Yan2019,ZY2021mixed,ZY2020improved,Yan2021,ZYan2022,yanEmergent2023} (shown in Fig.~\ref{fig:model}). The physics revealed therein~\cite{yanFully2022} is profound, such as the hidden vison plaquette (VP) crystal, which is invisible from dimer correlations and is sandwiched between the lattice nematic (LN) order and the $\mathbb{Z}_2$ quantum spin liquid (QSL) close to the Rokhsar-Kivelson (RK) point~\cite{Rokhsar1988,MoessnerSondhi2001b,henleyFrom2004}. Another interesting aspect is the structure of the phase diagram when connected to finite temperature, which is expected to be richer compared to its square lattice loop or dimer model cousins~\cite{Alet2005b,dabholkarReentrance2022,ranClassical2023}. However, perhaps the most intriguing aspect related to the quantum criticality of the model is the (2+1)D cubic* transition from the VP phase to the $\mathbb{Z}_2$ QSL. At this transition, the VP order parameter, which emerges from the underlying resonance of the dimer pairs, fractionalizes into the vison order parameter of the O(3)/cubic CFT primary field. The condensation of these fractionalized excitations, in turn, leads to a strong enhancement of the scaling dimension of the VP order parameter at the transition in an unconventional manner~\cite{YCWang2017QSL,GYSun2018,YCWang2018,Isakov2012,wangVestigial2021,YCWang2021NC}. 


Therefore, our motivation in this Letter is to elucidate the precise nature of the intriguing and unconventional cubic* quantum critical point (QCP) through both static and dynamic probes. We aim to achieve a comprehensive understanding of this QCP by combining field-theoretical interpretation with state-of-the-art QMC simulations. We find that this transition exhibits an anomalously large anomalous dimension when viewed through the correlation functions of the $t$-term (the resonance term in the QLM Hamiltonian, explained below). These correlation functions represent composite objects of the fractionalized vison and correspond to the rank-2 tensor (or tensorial magnetization) of the (2+1)D O(3)/cubic universality~\cite{Hasenbusch2023,hasenbusch2023phi4} with a large scaling dimension, approximately $\eta_T\approx1.42$~\cite{Ballesteros1996Finite,aharony1973critical,hasenbuschAnisotropic2011,Adzhemyan:2019gvv,Aharony:2022ajv,pelissettoCritical2002,Chester2021,rong2023o3}. On the other hand, if one measures the correlation of the vison operator, the observed anomalous dimension is consistent with the conventionally small values of $\eta\approx0.04$ for (2+1)D O(3)/cubic universality~\cite{Ballesteros1996Finite,aharony1973critical,hasenbuschAnisotropic2011,Adzhemyan:2019gvv,Aharony:2022ajv,pelissettoCritical2002,Chester2021}. This sharp contrast clearly reveals the unconventional nature of the cubic* transition that separates the unconventional VP phase, which is hidden from dimer measurements, from the $\mathbb{Z}_2$ QSL, where visons are the anyonic particles of the underlying topological order.

In addition to these purely theoretical motivations, the QLM that we studied here has been widely treated as the low-energy effective model for many frustrated magnets~\cite{Moessner2001l,moessnerShort2001,Krishanu2015,Plat2015,Yan2021,ZYan2022,yanEmergent2023,YCWang2017QSL,Wang2017,XFZhang2018,FengZL17,WenXG17,weiEvidence2017,liKosterlitz2020,huEvidence2020,wen2019experimental,Feng2018claringbullite,JJWen2019,YiZhou2017,Broholm2020,YDL2021,ZZ2020string,Moessner2001b,Ivanov2004,Ralko2005,Misguich2008,ranFully2023} and blockaded cold-atom arrays ~\cite{Samajdar:2020hsw,Verresen:2020dmk,zhou2022u1,Samajdar2022,Semeghini21} in condensed matter and cold atom experiments~\cite{Semeghini21,scholl2021quantum}.
In the Rydberg array, static characteristics can be easily obtained via the snapshot technique~\cite{scholl2021quantum, Semeghini21, tran2023measuring, wurtz2023aquila}, while dynamic information can be measured through real-time evolution~\cite{surace2020lattice, notarnicola2020real, guardadoProbing2018, turner2018weak}. Similarly, static and dynamic information for quantum magnets can be detected by neutron scattering or nuclear magnetic resonance experiments~\cite{FengZL17, weiEvidence2017, huEvidence2020, paddison2017continuous, zorkoSymmetry2017, wen2019experimental, zengDirac2023}, and our computational scheme of QMC + stochastic analytic continuation (SAC)~\cite{Sandvik1998a, Beach2004, Syljuasen2008, SHAO2023Progress} for the frustrated spin model, QDM, and QLM models has provided consistent static and dynamic information that has been used to explain experiments~\cite{Shao2017Nearly, liKosterlitz2020, huEvidence2020, GYSun2018, Ma2018a, YCWang2021NC, wangVestigial2021, CKZhou2021, Yan2021, ZYan2022, yanEmergent2023, bera2022emergent}. Based on these previous experiences, in this Letter, our QMC static correlations reveal different scaling dimensions at the cubic* QCP, corresponding to the different constituent operators in the CFT data for $\eta_T$ and $\eta$. At the same time, our QMC+SAC dynamic measurements exhibit continua of the dimer and vison spectra as the dynamic signature of the $\mathbb{Z}_2$ topological order and its associated vison condensation in the vicinity of the cubic* QCP.

\noindent{\textcolor{blue}{\it Model and Methods.}---} 
The Hamiltonian of the QLM on a triangular lattice is defined as
\begin{eqnarray}
  H=&-t&\sum_\alpha \left(
  \left|\rhombV\right>\left<\rhombH\right| + \mathrm{H.c.}
  \right) \nonumber \\
  &+V&\sum_\alpha\left(
  \left|\rhombV\right>\left<\rhombV\right|+\left|\rhombH\right>\left<\rhombH\right|
  \right),
\label{eq:eq1}
\end{eqnarray}
where $\alpha$ denotes all the rhombi (with three orientations) on the triangular lattice, as shown in Fig.~\ref{fig:model}(a). The local constraint of the fully packed QLM requires two dimers to touch every site in any configuration. The kinetic term is controlled by $t$, which generates dimer pair resonance on every flippable plaquette while respecting the local constraint, and $V$ is the repulsion ($V>0$) or attraction ($V<0$) between dimers facing each other. The RK point is located at $V=t=1$ and has an exact $\mathbb{Z}_2$ QSL solution~\cite{Moessner2001l}. We set $t=1$ as the energy unit and perform simulations for system sizes $L=8,12,16,20,24$ with the inverse temperature $\beta=\frac{1}{T}=L$ using the sweeping cluster QMC methods~\cite{Yan2019, ZY2020improved, Yan2021, ZYan2022, ZY2022}, and utilize the SAC scheme~\cite{Sandvik1998Stochastic, Shao2017Nearly, GYSun2018, Ma2018a, Yan2021, SHAO2023Progress, liKosterlitz2020, huEvidence2020, YCWang2021NC, CKZhou2021, ZYan2022} to obtain both the dimer and vison spectral functions in real frequency for $L=6,12$ systems from imaginary time correlation functions with $\tau \in [0, \beta=200]$.   

According to Ref.~\cite{yanFully2022}, the order parameter of the VP phase is given by the real space $t$-term correlation function
\begin{equation}
\label{eq:tpo}
\langle T(0)T(\mathbf{r}) \rangle=\frac{1}{3}[\langle t_{1}(0)t_{1}(\mathbf{r})\rangle+\langle t_{2}(0)t_{2}(\mathbf{r})\rangle+\langle t_{3}(0)t_{3}(\mathbf{r})\rangle],
\end{equation}
where $\langle t_{\alpha}(0)t_{\alpha}(\mathbf{r})\rangle$ $(\alpha=1,2,3)$ represent correlators on the three rhombus directions in our triangular lattice with distance $\mathbf{r}$ between two rhombi. 
The reason for discarding the off-diagonal terms in Eq.~\eqref{eq:tpo} will be explained below Eq.~\eqref{eq:eq2}.
The vison correlation function, constructed from the dimer configurations, is
\begin{equation}
    \langle \bar{v}(0)\bar{v}(\mathbf{r}) \rangle=\frac{1}{2}[\langle v_{1}(0)v_{1}(\mathbf{r})\rangle+\langle v_{2}(0)v_{2}(\mathbf{r})\rangle],
    \label{eq:eq3}
\end{equation}
where $v_\gamma$ $(\gamma=1,2)$ for the A (lower triangle) and B (upper triangle) sublattices in one rhombus. For the non-Bravais lattice, we only consider the diagonal terms of the correlation matrix $\langle\bar{v}_i(0)\bar{v}_j(r)\rangle$, and what we actually calculate is the trace of this matrix, i.e., $\text{Tr}(\langle \bar{v}_i(0)\bar{v}_j(r)\rangle)$. To obtain the vison configuration from dimer configuration, one needs to fix a gauge with the reference vison in the plaquette $(0,0)$ and sublattice A as $v_{1}(0)=1$, as shown in the schematic plot of Fig.~\ref{fig:model} (b). Then we map the dimer pattern to the vison configuration through $ v_{1}(0) v_{\gamma}(\mathbf{r}) $\,$=$\,$(-1)^{N_{P}}$, with $N_{P}$ being the number of dimers cut along the path $P$ between triangle at $0$ and $\mathbf{r}$, which refer to the green dashed line in the Fig.~\ref{fig:model} (b). Therefore, the vison in each triangle holds the value $\pm1$, as denoted by the red ($+1$) and grey ($-1$) triangles in the schematic plots of Fig.~\ref{fig:model} (b).  

\begin{figure}[htp!]
\centering
\includegraphics[width=1\columnwidth]{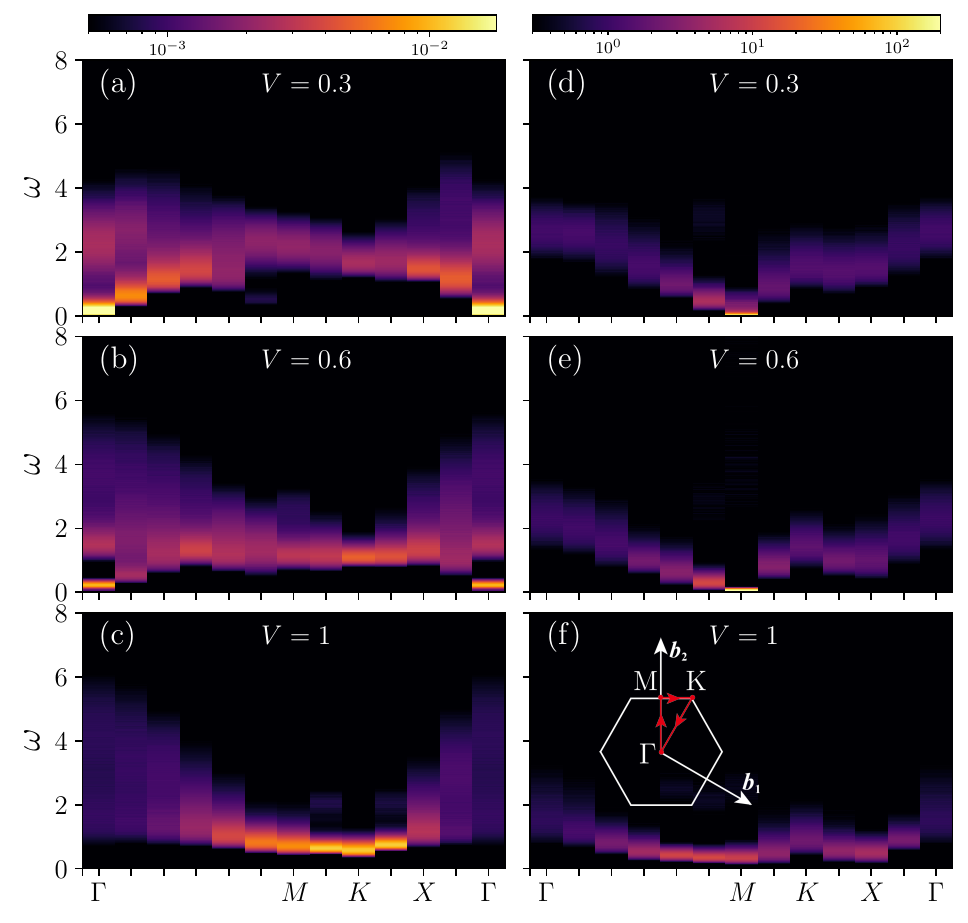}
\caption{{\textbf{The dynamic dimer and vison spectra.}} The spectra in the VP phase $(V=0.3)$, at the cubic* QCP $(V=0.6)$, and at the RK point $(V=1)$ for $L=12$ system. The $\beta$ used in the simulations is $100$ and we employ QMC+SAC scheme to generate the real frequency data. The inset in (f) shows the high-symmetry path in the Brillouin zone along which the spectra are presented. In the dimer spectra displayed in the left column, panels (a), (b), and (c), a gap is observed at the $\mathbf{M}$ point, suggesting that the dimer correlator cannot detect the transition between the VP and QSL phases. Conversely, the vison spectra in the right column, panels (d), (e), and (f), reveal a gap closure at the $\mathbf{M}$ point at the cubic* QCP in panel (e) and a reopening at the RK point in (f), which clearly indicates the VP-QSL transition.}
	\label{fig:spec}
\end{figure}

In the field theoretical description~\cite{yanFully2022}, the cubic* CFT of the VP-QSL transition can be described with three scalars coupled together. 
The Lagrangian is 
\begin{equation}
   \mathcal{L}_{int}=m^2 (\sum_i\phi_i^2)+ +u(\sum_i\phi_i^2)^2+v(\sum_i\phi_i^4)+\cdots,
\end{equation}
together with kinetic terms for the scalars, where the scalar order parameter describing the vison modes~\cite{Blankschtein1984,Blankschtein19842,Huh2011,Krishanu2015} is given by
\begin{equation}
\phi^{}_{j}=\sum_{\mathbf{r}}(v^{}_{1}(\mathbf{r}),v^{}_{2}(\mathbf{r}))\cdot\mathbf{u}^{}_{j}e^{i \mathbf{M}_j\cdot \mathbf{r}}, \quad j=1,2,3,
\label{eq:eq2}
\end{equation}
with $\mathbf{M}_{j=1,2,3}$ the three $\mathbf{M}$ points of the Brillouin zone as shown in the inset of Fig.~\ref{fig:spec} (b) and $v_{1,2}(\mathbf{r})$ the vison fields in Eq.~\eqref{eq:eq3}. The vector $\bm{\phi}$\,$=$\,$(\phi_{1},\phi_2,\phi_3)$ encapsulates the (2+1)D cubic order parameters of the visons.
The mass term can be roughly identified as $m^2\sim |V-V_c|$, and the phase transition happens at $m^2=0$. 
Conformal field theory tells us the correlation of $\phi$ fields follows a power law behavior. 
At the phase transition, the quantum fluctuation of the vison field is dominated by their modes at the $\mathbf{M}$ points. 
The vison correlation in Eq.~\eqref{eq:eq3} therefore will follow the same power law (with spatial modulation). 

As mentioned above, the $t$-term operator $t_i$ can be identified as the field theory operators, $\{t_1, t_2, t_3\}\sim \{\phi_1\phi_2, \phi_2\phi_3, -\phi_1\phi_3\}$. 
The symmetry group of the CFT is the cubic(3)=$S_3\rtimes(Z_2)^3$ group, the group elements of cubic(3) can be identified with lattice symmetries. 
The precise identification of $t$-operators is fixed by the symmetries that they break. Here we are following the convention of Ref.~\cite{yanFully2022}.
The cubic(3) group is a subgroup of O(3). It is known, based on various theoretical works \cite{Ballesteros1996Finite,Calabrese:2002sz,Chester2021}, that the O(3) CFT and the cubic(3) are connected by a very short renormalizations group flow, therefore their operators have similar anomalous dimensions. 
In particular, the O(3) group has a rank-2 symmetric traceless tensor representation, formed by $\{\phi_1\phi_2, \phi_2\phi_3, -\phi_1\phi_3\}$ and $\{\phi_1^2-\phi_2^2,\phi_2^2-\phi_3^2\}$, which is five-dimensional. In view of the subgroup cubic(3), the triple $\{\phi_1\phi_2, \phi_2\phi_3, -\phi_1\phi_3\}$ forms a three-dimensional irreducible representation of the cubic(3) group.
We can safely use the well-known value of the critical exponents $\eta_{T}\approx 1.42$ of O(3) CFT to approximate its value at the cubic CFT~\cite{Chester2021}. 
The subscript ``$T$'' reminds us that it corresponds to the rank-2 tensor of O(3). Interestingly, the off-diagonal correlator $\langle t_1(0) t_2(\vec{r})\rangle$ decays much faster than the diagonal ones $\langle t_1(0) t_1(\vec{r})\rangle$, which is also a CFT prediction and we show these results in Fig.S1 in the SM~\cite{suppl}. The anomalous dimension of scalar $\{\phi_1,\phi_2,\phi_3\}$ for O(3) CFT, i.e. the vison $v_{1,2}$ correlation in Eq.~\eqref{eq:eq3}, on the other hand, is of very small value $\eta\approx 0.04$~\cite{Ballesteros1996Finite,Chester2021}.

\begin{figure}[htp]
\centering
\includegraphics[width=1\columnwidth]{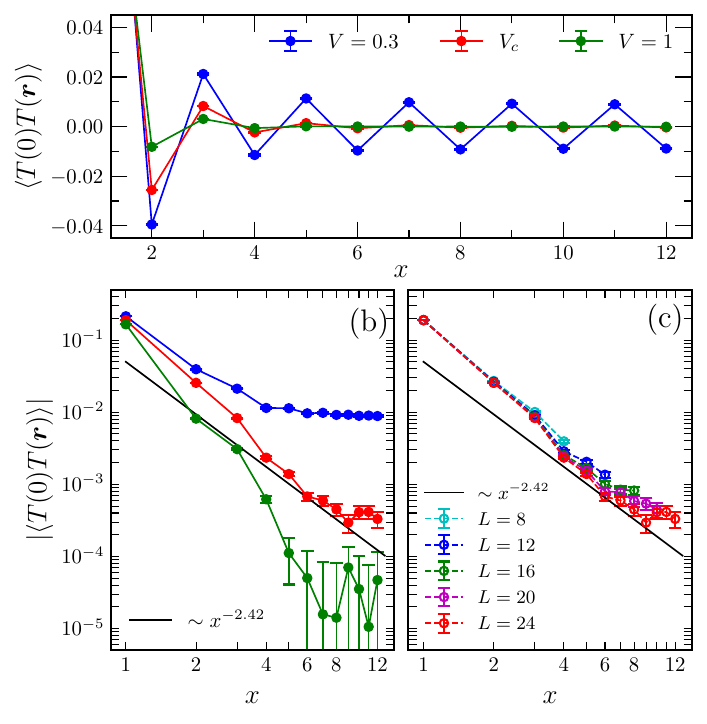}
\caption{{\textbf{The static $t$-term correlation.}} (a) $\langle T(0)T(\mathbf{r}) \rangle$ as a function of the distance $\mathbf{r}=(x,0)$ with the system size $L=24$, the largest system size achieved. The log-log plot for the absolute values of the data in (a) is shown in (b). We also show the log-log plot of the $t$-term correlators with different system sizes in (c) to demonstrate the finite-size effect of the decay behavior. (a), (b) The correlators in the VP phase with $V=0.3$, at the transition point with $V_c$ (we use $V=0.6$ here), and at the RK point when $V=1$. The dark solid lines shown in (b) and (c) are proportional to $1/x^{1+\eta_{T}}$ with the large anomalous dimension $\eta_{T}=1.42$, which corresponds to the rank-2 tensor field of the (2+1)D O(3)/cubic universality~\cite{Chester2021}.}
	\label{fig:tcor}
\end{figure}

\begin{figure}[htp]
\centering
\includegraphics[width=1\columnwidth]{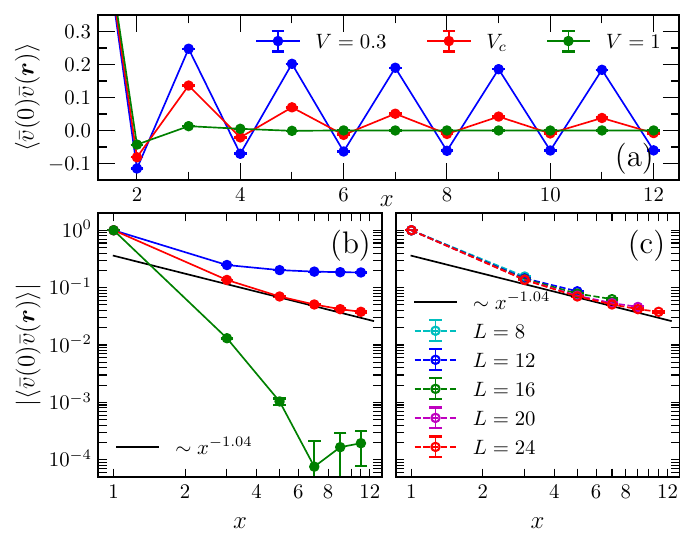}
\caption{{\textbf{The static vison correlation.}} (a) $\langle \bar{v}(0)\bar{v}(\mathbf{r}) \rangle$ as a function of the distance $\mathbf{r}=(x,0)$ with the fixed system size $L=24$. (b) The log-log plot of only the odd value of $x$ in (a). Similar to the $t$-term correlators, we show the vison correlators in the VP phase ($V=0.3$), at the transition point $V_c$ (use $V=0.61$ here), and at the RK point when $V=1$. (c) The outstanding critical decay behavior with different system sizes. Different from the $t$-term correlators in Figs.~\ref{fig:tcor} (b) and ~\ref{fig:tcor}(c), the dark solid lines shown in (b) and (c) here are proportional to $1/x^{1+\eta}$ with the anomalous dimension $\eta=0.04$ for the (2+1)D O(3)/cubic scalar order parameter~\cite{Ballesteros1996Finite,Chester2021}.}
\label{fig:vcor}
\end{figure}

We also compute the dynamic dimer correlation function 
$D(\mathbf{k},\tau)=\frac{1}{3N}
\sum_{\substack{i,j;\alpha=1,2,3}}^{L^2}
e^{i\mathbf{k}\cdot \mathbf{r}_{ij}} \left(\langle n^{}_{i,\alpha}(\tau)n^{}_{j,\alpha}(0)\rangle - \langle n^{}_{i,\alpha}\rangle \langle n^{}_{j,\alpha}\rangle \right)$,
where $n_{i,\alpha}$ is the dimer number operator on bond $i$ and $\alpha$ stands for the three bond orientations, and the vison dynamic correlation function
$\bar{v}(\mathbf{k},\tau)=\frac{1}{2N}
\sum_{\substack{i,j;\gamma=1,2}}^{L^2}
e^{i\mathbf{k}\cdot \mathbf{r}_{ij}} \left(\langle v^{}_{i,\gamma}(\tau)v^{}_{j,\gamma}(0)\rangle - \langle v^{}_{i,\gamma}\rangle \langle v^{}_{j,\gamma}\rangle \right)$,
which averages the correlation functions of visons in A and B sublattices. Since the value of vison in each triangle is $\pm1$, the second term $\langle v^{}_{i,\gamma}\rangle$ in $\bar{v}(\mathbf{k},\tau)$ is expected to be zero, i.e., no background needs to be subtracted.

\noindent{\textcolor{blue}{\it Numerical results.}---}
Figures.~\ref{fig:tcor} and \ref{fig:vcor} show the $\langle T(0)T(\mathbf{r}) \rangle$ and $\langle \bar{v}(0)\bar{v}(\mathbf{r}) \rangle$ across the cubic* QCP with system size up to $L=24$. The distance is along $\mathbf{r}=(x,0)$ with $x$ up to $12$ for the periodic boundary condition. The real-space decay behaviors is observed for both correlators in three regions: (i) VP phase with $V=0.3$. (ii) The cubic* QCP $V_c=0.59(2)$. (iii) The RK point $V=1$. 

In the VP phase, both $\langle T(0)T(\mathbf{r}) \rangle$ and $\langle \bar{v}(0)\bar{v}(\mathbf{r}) \rangle$ exhibit strong even-odd oscillations and with amplitude decaying with the distance $x$. The oscillations derive from the hidden vison order and eventually vanish as $V$ goes to 1 as shown in Figs.~\ref{fig:tcor} (a) and \ref{fig:vcor} (a). We note the even-odd oscillation still exists at the transition point due to the finite-size effect. The oscillations of all $V$ are symmetric with respect to $\langle T(0)T(\mathbf{r}) \rangle=0$, therefore, we illustrate $|\langle T(0)T(\mathbf{r}) \rangle|$ in log-log scale in Figs.~\ref{fig:tcor} (b) and ~\ref{fig:tcor}(c). Moreover, due to the gauge choice we set manually to construct the vison configuration, the oscillations of the vison correlation are asymmetrical with respect to $\langle \bar{v}(0)\bar{v}(\mathbf{r}) \rangle=0$ for different values of $V$. Thus, we only use the odd value of the distance to fit the data of $|\langle \bar{v}(0)\bar{v}(\mathbf{r}) \rangle|$ in log-log scale, as shown in Figs.~\ref{fig:vcor} (b) and ~\ref{fig:vcor}(c). 

We found in the VP phase both correlation functions decay to a constant value, while exhibiting power-law decay at the cubic* QCP. Interestingly, these two correlators decay with obvious different exponents. For the $t$-term correlation, $\langle T(0)T(\mathbf{r}) \rangle \sim 1/x^{1+\eta_{T}}$ is consistent with an anomalously large anomalous dimension of the rank-2 tensor of cubic CFT with $\eta_{T}=1.42$, and for the vison correlation $\langle \bar{v}(0)\bar{v}(\mathbf{r}) \rangle \sim 1/x^{1+\eta}$ is consistent with $\eta=0.04$, which is the (2+1)D O(3)/cubic value of $\eta$ for the order parameter. To access the thermodynamic limit, we depict correlators with different system sizes at the cubic* QCP in Figs.~\ref{fig:tcor} (c) and \ref{fig:vcor} (c), and put the small system sizes data of other values of $V$ in the SM~\cite{suppl}. All these results reveal the $\eta_T =1.42$ for $\langle T(0)T(\mathbf{r}) \rangle$ and the $\eta=0.04$ for the $\langle \bar{v}(0)\bar{v}(\mathbf{r}) \rangle$. On the other hand, inside the QSL phase such as $V=1$, the RK point, both correlators decay exponentially as shown in Figs.~\ref{fig:tcor} (b) and \ref{fig:vcor} (b).


Large anomalous dimension means a large scaling dimension as $\Delta_T=\frac{1+\eta_T}{2}$ for the rank-2 tensor and $\Delta=\frac{1+\eta}{2}$ for the scalar operators of the cubic/O(3) CFT, our results therefore mean that at the cubic* QCP, the $t_{1,2,3}$-term is a composite of the fractionalized visons $v_{1,2}$, instead of a well-defined critical mode, and it is the proliferated visons $v$ that  give rise to the large anomalous dimension of $t$, which serves as a defining signature of the cubic* transition, different from the conventional cubic/O(3) QCPs. Similar behavior has been observed in the (2+1)D XY* transition between the $\mathbb{Z}_2$ QSL and U(1) symmetry-breaking superfluid phase~\cite{YCWang2017QSL,Isakov2012,YCWang2018,YCWang2021NC}.

Such a fractionalization signature is also vividly seen from the dynamic probes. We measure the dynamic correlation functions $D(\mathbf{k},\tau)$ and $\bar{v}(\mathbf{k},\tau)$ and obtain the dimer and vison spectra via QMC+SAC (details of the scheme is given in the SM~\cite{suppl}). Figure~\ref{fig:spec} shows the obtained spectra across the cubic* transition. Inside the QSL phase denoted by Figs.~\ref{fig:spec} (c) and ~\ref{fig:spec}(f), both spectra exhibit gapped behavior and substantial continua in a large fraction of the momenta along the high-symmetry path. It is interesting to note that the minimal dimer gap is larger than the minimal vison gap due to the fact that a dimer is the composite of a pair of visons~\cite{feldnerDynamical2011,Yan2021}. 

At the cubic* QCP, the dimer spectra remain gapped at the $\mathbf{M}_{j=1,2,3}$ points, as shown in Fig.~\ref{fig:spec} (b). However, as depicted in Fig.~\ref{fig:spec} (e), the vison spectra develop a clear gapless mode close to the $\mathbf{M}$ points. Since the $\mathbf{M}$ points are the ordered wave vector of the VP phase [as explained in Eq.~\eqref{eq:eq2}], this critical and gapless vison mode serves as the dynamic signature of vison condensation at the cubic* transition. The contrast between Figs.~\ref{fig:spec} (b) and ~\ref{fig:spec}(e) explains why the dimer correlation cannot detect the "hidden" VP order, and only the vison spectra reveal the translational symmetry breaking of the VP phase. Similar dynamic signature of the $\mathbb{Z}_2$ topological order in QSL and the condensation of fractionalized anyons have also been demonstrated in the (2+1)D XY* transition~\cite{YCWang2017QSL,YCWang2018,GYSun2018,beckerDiagnosing2018,YCWang2021NC,wangVestigial2021}.

\noindent{\textcolor{blue}{\it Discussions.}---} Through a combined numerical and analytic approach, we have identified static and dynamic signatures of the cubic* transition from the $\mathbb{Z}_2$ QSL to the VP crystal in the QLM on a triangular lattice. Both correlations and spectra reveal that at the transition, the fractionalized vison in the QSL condenses, leading to the formation of the crystalline VP phase. This condensation leaves its trace in the anomalously large anomalous dimension exponent and pronounced continua in the dimer and vison spectra, distinguishing it from conventional cubic or O(3) quantum critical points. These findings reveal the underlying reason why the $t$-term correlation exactly corresponds to the rank-2 symmetric traceless tensor of the cubic/O(3) CFT and why the VP phase becomes invisible in dimer measurements. Moreover, we believe our findings will guide further experiments in frustrated quantum magnets and blocked cold-atom arrays, where the unconventional quantum matter and quantum phase transitions are being realized at an astonishing speed~\cite{FengZL17,WenXG17,weiEvidence2017,liKosterlitz2020,huEvidence2020,wen2019experimental,Feng2018claringbullite,JJWen2019,YiZhou2017,Broholm2020,YDL2021,ZZ2020string,Samajdar:2020hsw,Verresen:2020dmk,Semeghini21,ZYan2022,Samajdar2022,yanEmergent2023,zengDirac2023}.

{\noindent\it Acknowledgements---}We thank Ning Su, Fabien Alet, Rhine Samajdar and Subir Sachdev for insightful discussions on the O($3$) and cubic fixed points and the phase diagram of triangular lattice QLM. XXR, ZY and ZYM acknowledge the support from the Research Grants Council (RGC) of
Hong Kong Special Administrative Region of China (Project No. 17301721, No. AoE/P701/20, No. 17309822, No. C7037-22GF,
17302223), the ANR/RGC Joint Research Scheme sponsored by RGC of Hong Kong and French National Research Agency
(Project No. A\_HKU703/22). The K. C. Wong Education Foundation (Grant No. GJTD-2020-01). YQ acknowledges support from the the National Natural Science Foundation of China (Grant No. 11874115 and No. 12174068). JR is supported by Huawei Young Talents Program at IHES. Y.C.W. acknowledges  support from Zhejiang Provincial Natural Science Foundation of China (Grant No. LZ23A040003). We thank the Beĳng PARATERA Tech CO.,Ltd. (URL: https://cloud.paratera.com), the High performance Computing Centre of Beihang Hangzhou Innovation Institute Yuhang, HPC2021 system under the Information Technology Services and the Blackbody HPC system at the Department of Physics, University of Hong Kong for providing computational resources that have contributed to the research results in this paper.

\bibliographystyle{longapsrev4-2}
\bibliography{main}

\setcounter{equation}{0}
\setcounter{figure}{0}
\renewcommand{\theequation}{S\arabic{equation}}
\renewcommand{\thefigure}{S\arabic{figure}}

\clearpage
\newpage
\setcounter{page}{1}
\begin{widetext}
\linespread{1.05}
	
\centerline{\bf Supplemental Material for ``cubic* criticality emerging from quantum loop model on triangular lattice''} 
\vskip3mm

\centerline{}

In this Supplemental Material, we first present the average $t$-term correlation functions $\langle T(0)T(\mathbf{r}) \rangle$, the average vison correlations $\langle \bar{v}(0)\bar{v}(\mathbf{r}) \rangle$, and the the off-diagonal $t$-term correlations $\langle t_1(0)t_2(\mathbf{r}) \rangle$ with different system sizes in Fig.~\ref{fig:stv}. All this correlation functions, including the dynamic correlators which will be introduced later, are measured in the VP phase with $V=0.3$, at the transition point $V_c$, and at the RK point with $V=1$. Then, we discuss the details of the QMC+SAC scheme in Sec.~\ref{sec:sac}, including a brief introduction to the sweeping cluster QMC method. To demonstrate the SAC process, we utilize the dimer spectra as an example. The dimer and vison dynamic correlation functions as a function of the imaginary time $\tau$ for $k$ points
along the high-symmetry path are shown in Fig.~\ref{fig:gt}. The frequency dependence of the dimer and vison spectrum functions are presented in Fig.~\ref{fig:sw}. 


\section{The $t$-term and vison correlation functions}
\label{sec:cor}
In this section, we show the raw data for the average $t$-term and vison correlation functions with different system sizes in Fig.~\ref{fig:stv}. In the VP phase with $V=0.3$, both correlations exhibits strong even-odd oscillations and with amplitude decaying with the distance $x$ as shown in Fig.~\ref{fig:stv} (a) and (c). At the transition point, the oscillations become weak but not vanish due to the finite-size effect in Fig.~\ref{fig:stv} (g) and (i). At the RK point, which in the $\mathbb{Z}_{2}$ QSL phase, both correlations oscillate weakly along the $y=0$ horizontal line. 

We also measured the off-diagonal $t$-term correlations $\langle t_1(0)t_2(\mathbf{r}) \rangle$ as shown in the right hand side of Fig.~\ref{fig:stv}. The dark solid line shown in Fig.~\ref{fig:stv}(l) is proportional to $x^{-2.42}$, which indicates $\langle t_1(0)t_2(\mathbf{r}) \rangle$ decays faster than $\langle T(0)T(\mathbf{r}) \rangle$  at the cubic* transitions point.
\begin{figure}[htp]
	\centering
	\includegraphics[width=1\textwidth]{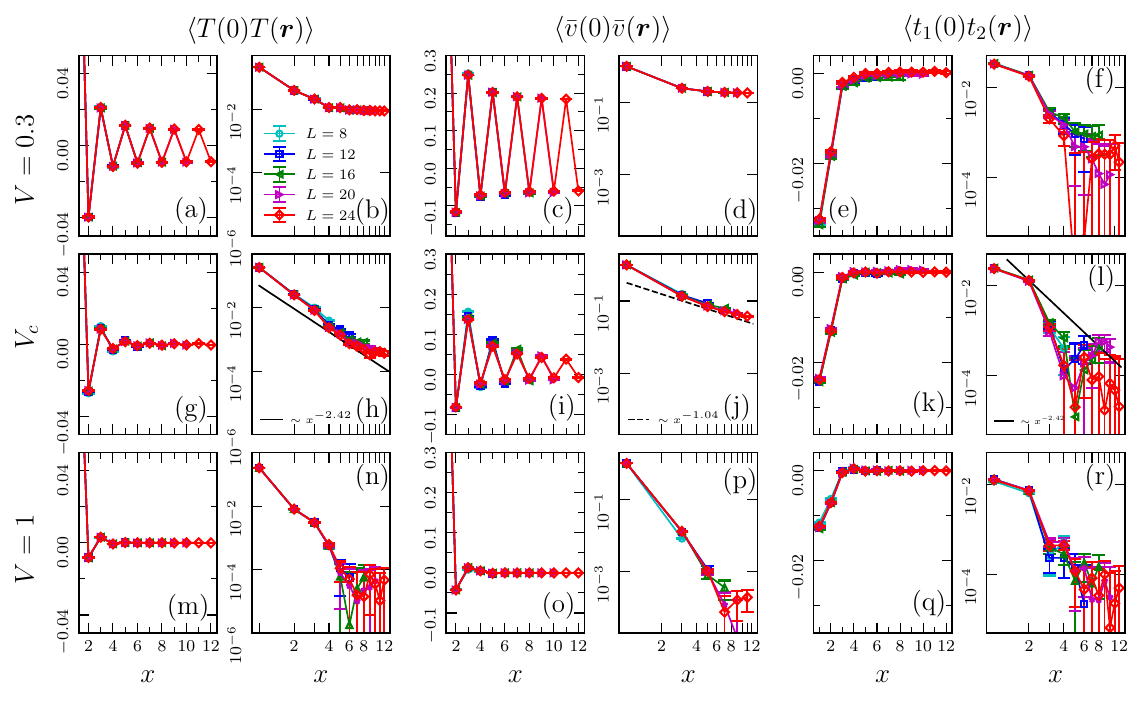}
	\caption{The diagonal $t$-term correlation function $\langle T(0)T(\mathbf{r}) \rangle$ (left), the diagonal vison correlation function $\langle \bar{v}(0)\bar{v}(\mathbf{r}) \rangle$ (middle), and the off-diagonal $t$-term correlation function $\langle t_1(0)t_2(\mathbf{r}) \rangle$ (right)  as a function of the distance $\mathbf{r}=(x,0)$ with different system sizes. We show all the correlators in the VP phase with $V=0.3$ in the upper row, at the transition point with $V_c$ in the middle row, and at the RK point when $V=1$ in the lower row. The log-log plots for the absolute values of the raw data are presented in the even column as well. For vison correlations, only odd $x$ data is presented for log-log plots in (d), (j) and (p). The dark solid lines shown in (h) and (l) are proportional to $x^{-2.42}$, and the dashed one in (i) is proportional to $x^{-1.04}$.}
	\label{fig:stv}
\end{figure}

\section{QMC+SAC SCHEME}
\label{sec:sac}

To obtain the dynamic spectral functions of the dimer and vison operators, we perform the SAC process to the dynamic correlations of the Eq.~\ref{eq:dt} and Eq.~\ref{eq:eq5} obtained in the QMC simulations. The sweeping cluster QMC approach employed in our work is a new method, whose main idea is to record the local constraints through sweeping and updating operators layer by layer along the imaginary time direction, thus all the samplings are performed in the restricted Hilbert space. This method is based on the world-line Monte Carlo scheme~\cite{OFS2002,Alet2005a,Alet2005b} and works well in constrained quantum lattice models~\cite{Yan2019,ZY2020improved,Yan2021,ZYan2022,yanFully2022}.

Here we take the dimer spectra as an example to demonstrate this QMC-SAC scheme. The dimer dynamic correlation functions $D(\mathbf{k},\tau)$, where $\mathbf{k}$ is the momentum point along the high-symmetry path as shown in Fig.~\ref{fig:spec}(c) in our simulations, relates to the real-frequency spectral function $S(\mathbf{k},\omega)$ through the equation
\begin{equation}\label{eqs1}
D(\mathbf{k},\tau)=\int_{-\infty}^{+\infty}S(\mathbf{k},\omega)e^{-\tau\omega}d\omega.
\end{equation}
With the relation between the positive and negative frequencies $S(\mathbf{k},-\omega)=e^{-\beta\omega}S(\mathbf{k},\omega)$, Eq.~\ref{eqs1} can be written only with positive frequencies
\begin{equation}\label{eqs2}
D(\mathbf{k},\tau)=\int_{0}^{\infty}K(\tau,\omega)S(\mathbf{k},\omega)D\omega,
\end{equation}
where the kernel is defined as
\begin{equation}
K(\tau,\omega)=\frac{1}{\pi}(e^{-\tau\omega}+e^{-(\beta-\tau)\omega}).
\end{equation}
Note that $D(\mathbf{k},\beta-\tau)=D(\mathbf{k},\tau)$ under this relation as well, thus we only need $\tau$ in the range of $[0,\beta/2]$. For the normalization $D(\mathbf{k},0)=1$, we therefore define $D(\mathbf{k},\omega)=(1+e^{-\beta\omega})S(\mathbf{k},\omega)/\pi$, where $\int_{0}^{\infty}D(\mathbf{k},\omega)D\omega=1$. Then, Eq.~\ref{eqs2} is replaced with
\begin{equation}\label{eqs4}
D(\mathbf{k},\tau)=\int_{0}^{\infty}\frac{e^{-\tau\omega}+e^{-(\beta-\tau)\omega}}{1+e^{-\beta\omega}}D(\mathbf{k},\omega)D\omega.
\end{equation}
 The main idea of the SAC is to obtain the estimate spectrum functions by means of the Monte Carlo sampling of the $\delta$-functions~\cite{Shao2017Nearly,SHAO2023Progress}. The weight function is $e^{-\chi^{2}/2\Theta}$, with $\chi^{2}$ the goodness of fit and $\Theta$ the sampling temperature. $\chi^2$ quantifies the difference between the average QMC data $\bar{D}(\mathbf{k},\tau)$ and the estimate correlations $D'(\mathbf{k},\tau)$, which is given by
 \begin{equation}
\chi^2=\sum_{i,j}(D'(\mathbf{k},\tau_i)-\bar{D}(\mathbf{k},\tau_i))C_{ij}^{-1}(D'(\mathbf{k},\tau_j)-\bar{D}(\mathbf{k},\tau_j)),
\end{equation}
which sum over all imaginary time $\tau_i, i=1,...,N_{\tau}$, a quadratic grid of $\tau_i$ is used with $\tau_i\propto i^2$\cite{SHAO2023Progress}. $C_{ij}$ is the covariance matrix element
 \begin{equation}
C_{ij}=\frac{1}{N_b(N_b-1)}\sum^{N_b}_{b=1}(D^{b}(\mathbf{k},\tau_i)-\bar{D}(\mathbf{k},\tau_i))(D^{b}(\mathbf{k},\tau_j)-\bar{D}(\mathbf{k},\tau_j)), 
\end{equation}
where $b=1,2,..,N_b$ denotes bins of the QMC data, $\bar{D}=\frac{1}{N_b}\sum_{b}D^{b}$. In the sampling process, we adjust $\Theta$ to find the minimal $\chi^{2}$ to satisfy $\langle \chi^2 \rangle\approx \chi^{2}_{min}+\sqrt{2N_{\tau}}$, which finally generate a smooth average spectral function.

Our QMC simulation results for the dynamic dimer and vison correlations as a function of the imaginary time $\tau$ for the fixed system size $L=12$ with $\beta=100$ are shown in Fig.~\ref{fig:gt}. $D(\mathbf{k},\tau)$ and $\bar{v}(\mathbf{k},\tau)$ with different high-symmetry momentum points are observed in the VP phase with $V=0.3$, at the VP-QSL transition point with $V=0.6$, and at the RK point where $V=1$. Besides, we also demonstrate the semi-log plots in the even column of Fig.~\ref{fig:gt} to show the exponential decay of the correlation data. For the dynamic dimer correlations, we found all $k$ points in the three phase exhibit the exponential decay with non-zero exponents, which indicates the spectra are all gapped in these momentum points. However, the decay exponents are very small when $V=0.3$ in Fig.~\ref{fig:gt}(h) and $V=0.6$ in Fig.~\ref{fig:gt}(j) for the dynamic vison correlations. 

To further confirm our prediction, the SAC method is performed to obtain the dimer spectra in the three phases. The frequency dependence of the dimer and vison spectrum functions $D(\mathbf{k},\omega)$ and $\bar{v}(\mathbf{k},\omega)$ are shown in the first and second row of Fig.~\ref{fig:sw} respectively, in which the peak of the $k$ point represents the strength of the dimer spectra in Fig.~\ref{fig:spec} in the main text. 
The gap at the $M$ point is a non-zero value in dimer spectra but seems very close to zero in viosn spectra. The spectrum at the $\Gamma$ point in the dimer spectra seems gapless may due to the finite-size effect.

\begin{figure}[htp!]
	\centering
	\includegraphics[width=1\textwidth]{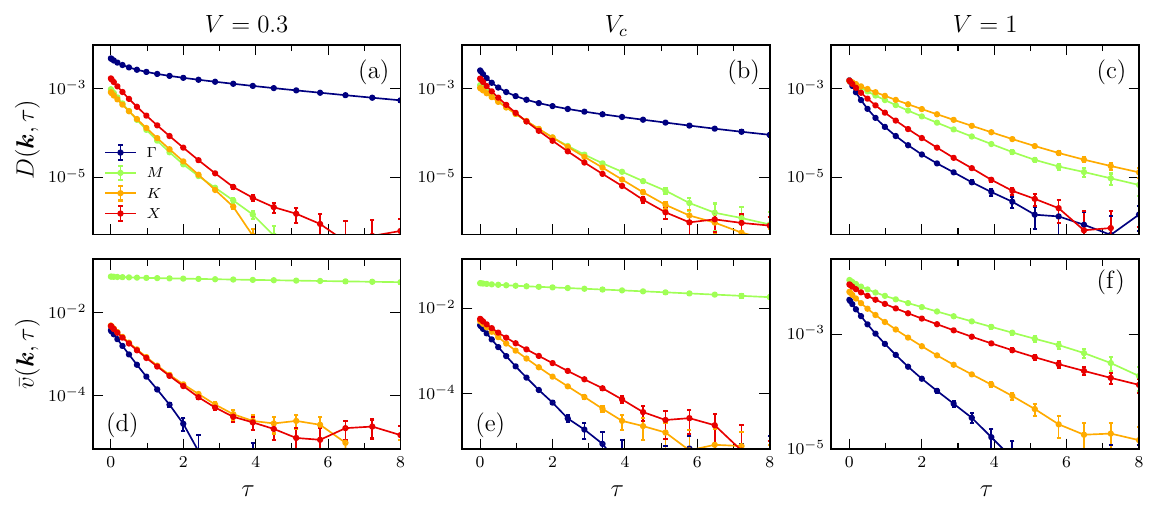}
	\caption{The semi-log plots of dimer (upper row) and vison (lower row) dynamic correlation functions obtained in QMC simulations with system size $L=12$ and $\beta=100$. Similar to the real-time correlation functions, $D(\mathbf{k},\tau)$ and $\bar{v}(\mathbf{k},\tau)$ are measured in the VP phase with $V=0.3$ (left), at the VP-QSL transition point with $V=0.6$ (middle), and at the RK point where $V=1$ (right). Selected $k$ points along the high-symmetry path shown in Fig.~\ref{fig:spec}(c) in the main text are presented to make the figures more clear.}
	\label{fig:gt}
\end{figure}

\begin{figure}[htp!]
	\centering
	\includegraphics[width=1\textwidth]{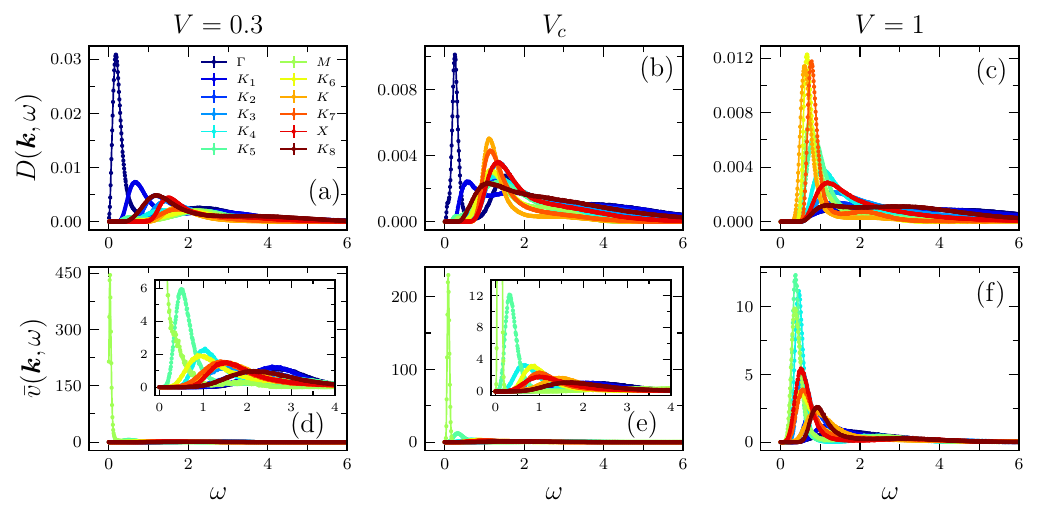}
	\caption{The dimer spectrum functions $D(\mathbf{k},\omega)$ and the vison spectrum functions $\bar{v}(\mathbf{k},\omega)$ as a function of frequency, which generated through SAC process from dynamic dimer and vison correlations data obtained from QMC,  are shown in the first and second row respectively. $k$ points along the high-symmetry path are illustrated in the VP phase with $V=0.3$ in (a) and (d), at the VP-QSL transition point with $V=0.6$ in (b) and (e), and at the RK point where $V=1$ in (c) and (f). The peaks of each $k$ point corresponding to the strength of the spectra in Fig.~\ref{fig:spec}. Insets in (d) and (e) zoom in the data to display $k$ points with small peaks. Note that the maximum values of the spectrum functions show different magnitudes for dimer and vison spectra, thus we have used multi colorbars for demonstration in Fig.~\ref{fig:spec}.     }
	\label{fig:sw}
\end{figure}


\end{widetext}

\end{document}